\documentstyle[12pt]{article}
\begin{document}
\begin{titlepage}

\begin{flushright}
ITP-UH-08/95
\end{flushright}

\begin{center}
{\bf\Large QUANTIZATION OF HIGHER SPIN SUPERFIELDS \\
IN THE ANTI--DE SITTER SUPERSPACE}
\end{center}
\vspace{1.5cm}

\noindent
{\large I.L. Buchbinder\\
{\it Department of Theoretical Physics\\
Tomsk State Pedagogical Institute, Tomsk 634041, Russia\\}}
\vspace{0.6cm}

\noindent
{\large S.M. Kuzenko}${}^{\dag\ddag}$\footnote{Alexander von
Humboldt research fellow}\\
${}^{\dag} ${\large\it Institut f\"{u}r Theoretische Physik,
Universit\"{a}t Hannover\\ 30167 Hannover, Germany}

\setcounter{footnote}{0}
\vspace{0.4cm}

\noindent
and

\vspace{0.4cm}

\noindent
{\large A.G. Sibiryakov}\\
${}^{\ddag}${\large\it Department of Quantum Field Theory\\
Tomsk State University, Tomsk 634050, Russia}

\vspace{1.5cm}

\begin{abstract}
Lagrangian quantization of the free superfield gauge theories of higher
massless superspins is performed both in the anti-de Sitter and flat
superspaces. For the models under consideration the straightforward
application of the BV procedure for quantization of reducible theories leads
to immense calculations. In order to avoid the difficulty, a simplification
of this procedure in reduction coordinates is considered. The contribution to
the effective action is shown to be independent on the gauge structure of the
classical formulations dually equivalent to each other. It is the same both
for the actions with finite and infinite stages of reducibility.
\end{abstract}
\vfill
\null
\end{titlepage}

\newpage

\noindent
{\bf 1.} Recently superfield gauge formulations have been constructed
in the $N=1$, $D=4$ flat superspace for arbitrary massless multiplets
of higher superspins [1,2], and their anti-de Sitter (AdS)
counterparts have been found [3]. Thus the old problem of finding
off-shell superfield realizations for the massless unitary
representations of the Poincar\'e superalgebra has been completely
solved and the results have been naturally extended in the AdS
superspace. Previously such realizations were known only for the
multiplets of lower superspins $s\le3/2$ where the choice $s=3/2$
corresponds to linearized supergravity. The actions obtained in
Refs.[1--3] constitute the manifestly supersymmetric formulations of
the known free higher-spin theories given in the papers [4--8] in the
form with an implicit supersymmetry [6,9,10]. The formulations of
Ref.[3] are explicitly invariant under the action of the AdS
superalgebra $osp(1,4)$ and realize on-shell massless higher-spin
representations of this superalgebra.

Some years ago there was a considerable progress [10--14] towards the
solution of the famous higher-spin problem. In particular it was
shown that consistent gravitational interactions of massless
higher-spin fields exist, at least in the first nontrivial order, but
turn out to be non-analytic in the cosmological constant. In this
respect the AdS formulations of Ref.[3] represent the necessary step
for development of a superfield approach to the higher-spin problem.

It was mentioned in Ref.[3] that the obtained gauge models are
reducible according to the terminology of Lagrangian quantization
[15]. In the flat superspace the stage of reducibility is finite or
infinite depending on the superspin and the type of the formulation.
It means that there exists an analogy between some of the obtained
models and the Green-Schwartz superstring theory. The remarkable
feature of the formulations in the AdS superspace is that there is
always a covariant replacement of gauge parameters which converts the
infinite stage of redusibility to the finite one. This arises
interest in the problem of quantization. For the superfield
supergravity (the first stage of reducibility) the question of
quantization was discussed in Refs.[16,17].

Quantization of the theories under consideration opens a possibility
to set up the problem of calculating effective action.	The effective
action corresponding to lower-spin fields in the AdS space was
investigated in a number of papers (see e.g. [18--25]). As to
higher-spin field contributions to effective action in AdS space, they
have not been considered so far. Clearly this problem is closely
related to that of the higher-spin field propagators in the AdS space
(see Ref.[25] where such a problem was studied from the
supersymmetric point of view). We expect that our research allows to
develop a completely superfield approach to the problem of effective
action induced by higher-spin superfields in the AdS
superspace\footnote{Completely superfield approach to effective
action of lower spin superfields has been developed in early papers
[28,29].}.

In the present letter we perform the Lagrangian quantization for the
formulations of Refs.[1--3]. To investigate the question the BV method [15]
can be used, but its straightforward application for these models leads to
immense nonlocal calculations.	In order to avoid this difficulty we describe
some special simplification of the BV procedure in reduction coordinates for
a general quadratic action. In our case these coordinates are given by
transversal irreducible superfields (ISes). The path integrals over the space
of such superfields are expressed in terms of those over unconstrained
superfields and chiral scalars. It allows us to receive a neat natural result
and explain why in the case at hand the infinite stage of reducibility does
not lead to obstacles in quantization.

\vspace{0.5cm}
\newcommand{\D}{{\cal D}}				 %new
\newcommand{\DB}{\bar{\cal D}}				 %new
\newcommand{\A}{\alpha} 				 %new
\newcommand{\AD}{\dot{\alpha}}				 %new
\newcommand{\BD}{\dot{\beta}}				 %new
\newcommand{\B}{\beta}					 %new
\noindent
{\bf 2.} The key feature of the formulations of Ref.[3] is the use
of transversally and longitudinally linear superfields. A complex
symmetric superfield $\Gamma (s-1,s-1)$\footnote{Our notations coincide
mainly with those adopted in [26,27], in particular
$\D_A=(\D_a,\D_\A,\DB^{\AD})$
are the super AdS covariant derivatives. All superfields in this
letter are symmetric in their undotted and dotted indices separately,
the number of indices being just indicated in parentheses:
$\Psi (k,l) \equiv \Psi _{\A(k)\AD(l)}	 \equiv
\Psi_{\A_1 \ldots \A_k\AD_1\ldots \AD_l}
\equiv \Psi_{(\A_1 \ldots \A_k)(\AD_1\ldots \AD_l)}$.}	satisfying
the constraint
\begin{eqnarray}
\DB^{\dot{\beta}}
\Gamma_{\alpha (s-1)\dot{\beta}\dot{\alpha}(s-2)}=0
&\Leftrightarrow&
(\DB^2-2(s+1)\mu )
\Gamma_{\alpha (s-1)\dot{\alpha}(s-1)}=0
\end{eqnarray}
is called transversally linear. A complex symmetric superfield
$\Gamma (s-1,s-1)$ satisfying the constraint
\begin{eqnarray}
\bar{\cal D}_{(\dot{\alpha}_1}
G_{\alpha (s-1)\dot{\alpha}(s-1))}=0
&\Leftrightarrow&
(\bar{\cal D}^2+2(s-1)\mu )
G_{\alpha (s-1)\dot{\alpha}(s-1)}=0
\end{eqnarray}
(the symmetrization all over the dotted indices is indicated) is called
longitudinally linear. In quantization superfields $\Gamma$, $G$ are to be
expressed in terms of unconstrained superfields $\Psi (s-1,s)$ and $\Psi
(s-1,s-2)$ by the rule
\begin{eqnarray} \Gamma_{\alpha
(s-1)\dot{\alpha}(s-1)}&=& \bar{\cal D}^{\dot{\beta}}
\Psi_{\alpha(s-1)\dot{\beta}\dot{\alpha}(s-1)}		  \\
G_{\alpha (s-1)\dot{\alpha}(s-1)}&=&
\DB_{(\dot{\alpha}}
\Psi_{\alpha (s-1)\dot{\alpha}(s-2))}.			  % (4)
\end{eqnarray}
Being expressed in these terms the actions of the transversal and longitudinal
formulations of half-integer superspin $s+1/2$ look like
\newcommand{\HU}{H^{\alpha(s)\dot{\alpha}(s)}}	     %new
\newcommand{\HD}{H_{\alpha(s)\dot{\alpha}(s)}}	     %new
\begin{eqnarray}
&S_{s+1/2}^{\bot} =
(-\frac{1}{2})^s \int d^8z \, E^{-1} \Big\{ \frac{1}{8}
\HU \D^{\beta} (\DB^2-4\mu) \D_{\beta} \HD	 &\nonumber\\
&\mbox{}-\frac{s^2}{2}\mu\bar{\mu}\HU \HD +
[\HU \D_{\alpha_1}\DB_{\dot{\alpha}_1}\DB^{\dot{\beta}}
\Psi_{\alpha(s-1)\dot{\beta}\dot{\alpha}(s-1)}	       &     \\
&\mbox{}-\frac{s+1}{s}\Psi^{\alpha(s-1)\dot{\beta}\dot{\alpha}(s-1)}
\DB_{\dot{\beta}}\DB^{\dot{\gamma}}
\Psi_{\alpha(s-1)\dot{\gamma}\dot{\alpha}(s-1)} + c.c.] +
2\Psi^{\alpha(s-1)\dot{\alpha}(s)}
\DB_{\dot{\alpha}_1}\D^{\alpha_1}
\bar{\Psi}_{\alpha(s)\dot{\alpha}(s-1)}  \Big\}      &	   \nonumber
\end{eqnarray}
\begin{eqnarray}
&S_{s+1/2}^{\|} =
(-\frac{1}{2})^s\int d^8z\, E^{-1} \Big\{ \frac{1}{8}
\HU \D^{\B} (\DB^2-4\mu) \D_{\B} \HD	     & \nonumber\\
&\mbox{}-\frac{1}{8}\frac{s}{2s+1} [\D_{\B},\DB_{\BD}]
H^{\B\A(s-1)\BD\A(s-1)}
[\D^{\gamma}\DB^{\dot{\gamma}}]
H_{\gamma\A(s-1)\dot{\gamma}\AD(s-1)}	  &   \nonumber\\
&\mbox{}-\frac{s}{2} \D^{\gamma\dot{\gamma}}
H_{\gamma\A(s-1)\dot{\gamma}\AD(s-1)}
\D_{\B\BD}
H^{\B\A(s-1)\BD\AD(s-1)} +
\frac{s^2}{2} \mu\bar{\mu} \HU \HD	      &       \nonumber\\
&\mbox{}+[\frac{2is}{2s+1} \D_{\gamma\dot{\gamma}}
H^{\gamma\A(s-1)\dot{\gamma}\AD(s-1)}
\DB_{\AD_1} \Psi_{\A(s-1)\AD(s-2)}		&	       \\
&\mbox{}-\frac{s+1}{s(2s+1)} \Psi^{\A(s-1)\AD(s-2)}
\DB^{\AD_1}\DB_{(\AD_1}
\Psi_{\A(s-1)\AD(s-2) ) } + c.c. ]	  &	  \nonumber\\
&\mbox{}-\frac{2}{2s+1} \Psi^{\A(s-1)\AD(s-2)}
{\DB^{\AD_1}\D_{\A_1}}
\bar{\Psi}_{\A(s-2)\AD(s-1)} \Big\}	       &       \nonumber
\end{eqnarray}
Here $d^8z\, E^{-1}$ is the super AdS invariant measure, $\mu$ is the
curvature of the AdS superspace,
$\DB^2=\DB_{\dot{\alpha}} \DB^{\dot{\alpha}}$ and c.c.
stands for complex conjugation.

The gauge structure of the actions (5,6) is as follows
\begin{eqnarray}
&\delta \HD = \DB_{(\dot{\alpha}_1} L^0_{\alpha(s)\dot{\alpha}(s-1))} -
\D_{(\alpha_1}
\bar{L}^0_{\alpha(s-1))\dot{\alpha}(s)} 		 &
\end{eqnarray}
\begin{eqnarray}
&\delta L^k_{\alpha(s)\dot{\alpha}(s-k-1)} =
\DB_{(\dot{\alpha}_1} L^{k-1}_{\alpha(s)\dot{\alpha}(s-k-2))},
\; k=0,\ldots,s-2					 &   \\
&\delta L^{s-1}_{\alpha(s)} = L^{s}_{\alpha(s)}, \;\;\;\;
\; \DB_{\dot{\alpha}} L^{s}_{\alpha(s)}=0		  & %(9)
\end{eqnarray}
\begin{eqnarray}
&\delta \Psi_{\alpha(s-1)\dot{\alpha}(s)} =
-\frac{s}{2(s+1)} \D^{\beta}\D_{(\beta}
\bar{L}^0_{\alpha(s-1))\dot{\alpha}(s)} +
\DB^{\dot{\beta}} \epsilon^0_{\alpha(s-1)\dot{\alpha}(s)\dot{\beta}}&
\end{eqnarray}
\begin{eqnarray}
&\delta \epsilon^k_{\alpha(s-1)\dot{\alpha}(s+k+1)} =
\DB^{\BD} \epsilon^{k+1}_{\A(s-1)\AD(s+k+1)\BD}, \;\;
 k=0,1,\ldots,\infty					       &%(11)
\end{eqnarray}
\begin{eqnarray}
&\delta \Psi_{\alpha(s-1)\dot{\alpha}(s-2)} =
-\frac{1}{2} \DB^{\BD}\D^{\B} L^0_{\B\A(s-1)\BD\AD(s-2)} +
i(s-1)\D^{\B\BD} L^0_{\B\A(s-1)\BD\AD(s-2)}	  &    \nonumber\\
&\mbox{}+\DB_{(\AD_1} \epsilon^0_{\A(s-1)\AD(s-3))}		   & % (12)
\end{eqnarray}
\begin{eqnarray}
&\delta \epsilon^k_{\A(s-1)\AD(s-k-3)} = (-1)^k\frac{s}{s-k-1}
(\frac{1}{2}\DB^{\BD}\D^{\B} L^{k+1}_{\B\A(s-1)\BD\AD(s-k-3)}  &\nonumber\\
&\mbox{}-i(s-k-2)\D^{\B\BD} L^{k+1}_{\B\A(s-1)\BD\AD(s-k-3)}) +
\DB_{(\AD} \epsilon^{k+1}_{\A(s-1)\AD(s-k-4))}, \;\; k=0,\ldots,s-4 &\\%(13)
&\delta \epsilon^{s-3}_{\A(s-1)} = (-1)^{s-3}\frac{s}{2}(\frac{1}{2}
\DB^{\BD}\D^{\B} L^{s-2}_{\B\A(s-1)\BD} -
i\D^{\B\BD} L^{s-2}_{\B\A(s-1)\BD} ) + \epsilon^{s-2}_{\A(s-1)}, \;\;\;
\DB_{\AD} \epsilon^{s-2}_{\A(s-1)}=0				&  \\%(14)
&\delta \epsilon^{s-2}_{\A(s-1)} = (-)^s\frac{s}{4}
(\DB^2-4\mu) \D^{\B} L^{s-1}_{\B\A(s-1)}.		  &
\end{eqnarray}
Here $ L^k(s,s-k-1), k=0,\ldots,s-1$; $\epsilon^k(s-1,s+k+1),
k=0,\ldots,\infty$; $\epsilon^k(s-1,s-k-3), k=0,\ldots,s-3 $ are  complex
gauge parameters and $	 L^s_{\A(s)}$, $\epsilon^{s-2}_{\A(s-1)}   $   are
chiral	superfields.  Eqs.(5--15) give	 the   initial	 data	for   the
quantization   of  the formulations under  consideration.  The	 gauge
structure  (7--11)  of	 the transversal formulation   (5)   has    infinite
stage	of    redusibility while  the gauge  structure	(7--9,12--15)  of the
longitudinal formulation (6) (dually   equivalent   to	 the   former [3])
has  finite  stage  of redusibility $s$ [15].

We remark that both the formulations of the theories of integer superspin
have infinite chains of reducibility transformations (similar to (11))
already for the parameters $L^k$ instead of (8). There exists a possibility
in the AdS superspace to convert the infinite stage theory to the finite
stage one by the change of gauge parameters [3]. But this operation has no
correct flat limit.

\vspace{0.5cm}
\noindent
{\bf 3.} In quantization of the formulations (5,6) the main technical
difficulty is to bring the operator in the resultant action in path integral
to a diagonal form. Already in the case of the supergravity $(s=3/2)$
this requires the use of a nonlocal gauge [16] which for higher $s$
turns into that containing finite series in powers of $\Box^{-1}$ up to the
$s$th order.  Then so-called catalist ghosts are to be introduced [16]. For
higher $s$ each catalist requires several catalists of their own and so on up
to $s$th generation, that leads to an extremely formidable construction. Here
we suggest the way out of this difficulty.

Consider a quadratic action and its reducible gauge structure (in this
section we shall consider all the dynamical variables $\phi^i$ of the theory
to be bosonic)
\begin{eqnarray}
&s[\phi] = s_{ij}\phi^i\phi^j, \;\;\;& i=1,\ldots,n		 \\%(16)
&s_{ij} Z^j_{\A_0} = 0, \;\;\;& \A_0 = 1,\ldots,m_0		   \\%(17)
&Z^{\A_{k-1}}_{\A_k}Z_{\A_{k+1}}^{\A_k} = 0, \;\;\;&
\A_k =1,\ldots,m_k;\A_{-1}\equiv i.				%(18)
\end{eqnarray}
Let us split each index into two subsets by the rule
\begin{eqnarray}
&i = \{a,\mu_0\}, \;\;& a=1,\ldots,{\rm rank}\,s_{ij}		\\%(19)
&\A_k=\{\mu_k,\mu_{k+1}\}, \;\;&
\mu_k=1,\ldots,{\rm rank} \, Z_{\A_k}^{\A_{k-1}}		 %(20)
\end{eqnarray}
i.e. $a$ is running through the number of true physical degrees of freedom,
$\mu_0$ is running through the number of true gauge invariances and so on.
Now consider the homogeneous tensor transformation
\begin{eqnarray}
&\phi^i = V_a^i \phi^a + V_{\mu_0}^i \phi^{\mu_0}	    & \\ %(21)
&c_k^{\A_k} = V_{k \, \mu_k}^{\A_k} c_k^{\mu_k} +
V_{k \, \mu_{k+1}}^{\A_k} c_k^{\mu_{k+1}}		   &	 % (22)
\end{eqnarray}
satisfying the conditions
\begin{eqnarray}
&s_{ij}V^i_{\mu_0}=0,	\;\;\;\; Z_{\A_{k}}^{\A_{k-1}} V^{\A_k}_{k \,
\A_{k+1}} = 0.&
\end{eqnarray}
Then without going into the details we state that in the new coordinates from
the r.h.s. of (21,22) the whole BV action is given effectively by the
following
\begin{eqnarray} &S = \phi^a
\tilde{s}_{ab} \phi^b + \sum\limits_k \bar{c}_{k\mu_k}
\tilde{Z}_{\nu_k}^{\mu_k} c_k^{\nu_k}			   &	    %(24)
\end{eqnarray}
where $ \bar{c}_k, c_k $ are the Faddeev-Popov ghosts and their analogues for
higher stages and
\begin{eqnarray} &\tilde{s}_{ab} = V_a^i s_{ij} V_b^i, \;\;\;\;\;\;\;
\tilde{Z}_{\nu_k}^{\mu_k} = \left( V_{k-1}^{-1} \right)_{\A_{k-1}}^{\mu_k}
Z_{\A_k}^{\A_{k-1}} V^{\A_k}_{k \, \nu_k}	&\nonumber\\
&V_k^{-1} =
\left( V^{\A_k}_{k \, \mu_k}, V^{\A_k}_{k \, \mu_{k+1}} \right)^{-1}.& %(25)
\end{eqnarray}
When passing to the action (24) in path integral, the Jacobian of the
replacement (21,22) has to be accounted. It is important to note here
that at the $k$th stage except the ghost $ c_k^{\A_k} $ of the minimal
sector all the fields were introduced in [15] by pairs ghost +
Lagrangian multiplier with the opposite statistic. Hence the Jacobians
for every pair cancel and only one Jacobian of the change (22)
remains at the fixed stage.

\vspace{0.5cm}
\noindent
{\bf 4.} The key point in our construction is the use of some off-shell
irreducible superfields (ISes) to construct the decomposition (21,22) for the
quantization of the longitudinal formulation (6--9,12--15). We use the
abbreviation ''IS'' to name the transversally linear and antilinear
simultaneously superfield $ \zeta(k,l) $
\begin{eqnarray}
&(\DB^2-2(l+2)\mu) \zeta(k,l) = (\D^2-2(k+2)\mu) \zeta(k,l) = 0     &  % (26)
\end{eqnarray}
(see also (1)), and chiral superfield $ \sigma_{\A(k)} $
\begin{eqnarray}
\DB_{\AD} \sigma_{\A(k)}& =& 0. 				   % (27)
\end{eqnarray}
In the case $ k=l $ the additional reality condition is admissible
(such a real IS we shall denote  by $\rho$)
\begin{eqnarray}
&  \rho(k,k) = \bar{\rho}(k,k)			       &	% (28)
\end{eqnarray}
The superfields (26,27) describe irreducible complex representations
of the super AdS algebra upon specifying the value of the quadratic
Casimir
\begin{eqnarray}
Q = -\frac 12 \D_{\A\AD} \D^{\A\AD} +
\frac 14 (\mu \D^2 + \bar{\mu} \DB^2) - \mu\bar{\mu}
(M^{\A\B}M_{\A\B} + \bar{M}_{\AD\BD}\bar{M}^{\AD\BD} ),&&
[Q,\D_{\A}] = 0.					 % (29)
\end{eqnarray}

Now we define a parametrization of the superfields of the theory
(6--9,12--15), separating explicitly transversal and longitudinal parts
of the complex ones:
\begin{eqnarray}
&\HD = \sum\limits^s_{k=0} \D_{(\A_1(\AD_1}\ldots\D_{\A_{s-k}\AD_{s-k}}
\rho_{\A_{s-k+1}\ldots\A_s) \AD_{s-k+1}\ldots\AD_s)}  & \nonumber\\
&+ \sum\limits^{s-1}_{k=0} [\D_{(\A_1},\DB_{(\AD_1}]
\D_{\A_2\AD_2}\ldots\D_{\A_{s-k}\AD_{s-k}}
\rho'_{\A_{s-k+1}\ldots\A_s) \AD_{s-k+1}\ldots\AD_s)}   & \nonumber\\
&\mbox{} + \sum\limits^s_{k=1}
\D_{(\A_1(\AD_1}\ldots\D_{\A_{s-k}\AD_{s-k}} \{\DB_{\AD_{s-k+1}}
\zeta'_{\A_{s-k+1}\ldots\A_s) \AD_{s-k+1}\ldots\AD_s)}+c.c.\} &     \\%(30)
&+ \D_{(\A_1(\AD_1}\ldots\D_{\A_s)\AD_s)}
\{\sigma'+\bar{\sigma}'\} &\nonumber
\end{eqnarray}
\setcounter{equation}{0}
\renewcommand{\theequation}{31.\alph{equation}}
\begin{eqnarray}
&\Psi(s-1,s-2) = \Psi^\bot(s-1,s-2)+\Psi^\|(s-1,s-2)	&      % (31.a)
\end{eqnarray}
where
\begin{eqnarray}
&\Psi^\bot_{\A(s-1)\AD(s-2)} = \sum\limits^{s-2}_{k=0}
\DB^{\BD}\D_{(\A_1(\BD}\D_{\A_2\AD_1}\ldots \D_{\A_{s-k-1}\AD_{s-k-2}}
\zeta_{\A_{s-k}\ldots\A_{s-1}) \AD_{s-k-1}\ldots\AD_{s-2})}  &\nonumber\\
&+\sum\limits^{s-1}_{k=1} \frac 1{\mu} (\DB^2+2(s-2)\mu)
\D_{(\A_1(\AD_1}\ldots \D_{\A_{s-k-1}\AD_{s-k-1}}
\zeta_{\A_{s-k}\ldots\A_{s-1})
\AD_{s-k}\ldots\AD_{s-2})}			&		 \\ % (31.b)
&+\D_{(\A_1}\D_{\A_2(\AD_1}\ldots\D_{\A_{s-1})\AD_{s-2})}\sigma& \nonumber
\end{eqnarray}
\begin{eqnarray}
&\Psi^\|_{\A(s-1)\AD(s-2)} = \sum\limits^{s-2}_{k=1}
\DB_{(\AD_1}\D_{(\A_1\AD_2}\ldots \D_{\A_{s-k-2}\AD_{s-k-1}}
\zeta''_{\A_{s-k-1}\ldots\A_{s-1}) \AD_{s-k}\ldots\AD_{s-2})}&\nonumber\\
&+\sum\limits^{s-2}_{k=1} \frac 1{\mu} (\DB^2-2s\mu)
\D_{(\A_1(\AD_1}\ldots \D_{\A_{s-k-1}\AD_{s-k-1}}
\zeta''_{\A_{s-k}\ldots\A_{s-1})
\AD_{s-k}\ldots\AD_{s-2})}				   &	\\ % (31.c)
&+\D_{(\A_1(\AD_1}\ldots\D_{\A_{s-2}\AD_{s-2})}
\sigma''_{\A_{s-1})}                                & \nonumber
\end{eqnarray}
\setcounter{equation}{31}
\renewcommand{\theequation}{\arabic{equation}}
and similarly in more condense notations
\begin{eqnarray}
&L^p(s,s-p-1)
= L^{p\bot} (z^p(j+1,j-p),z^p(j,j-p),\phi^p_{\A(p)}) &\nonumber\\
&+L^{p\|} (m^p(k-1,k-p-1),m^p(k,k-p-1),\mu^p_{\A(p+1)})   &	 \\%(32)
& j=p,\ldots,s-1, \;\;\;\; k=p+1,\ldots,s-1		     &\nonumber
\end{eqnarray}
\begin{eqnarray}
&\epsilon^p(s-1,s-p-3) =
\epsilon^{p\bot} (n^p(j+1,j-p-1),n^p(j,j-p-1),\nu^p_{\A(p+1)}) &\nonumber\\
&+\epsilon^{p\|} (a^p(k+1,k-p-2),a^p(k,k-p-2),\A^p_{\A(p+2)}) &   \\%(33)
&j=p+1,\ldots,s-2, \;\;\;\; k=p+2,\ldots,s-2.		      &\nonumber
\end{eqnarray}
In Eqs.(30-33) $ \zeta$, $z$, $m$, $n$ and $a$ are subjected to the
conditions (26), $\rho$ satisfies (26,28) and $ \sigma$, $\phi$, $\mu$,
$\nu$, $\A $ are chiral (27). Note that, owing to the constraints (1,26),
ISes with different total number of indices cannot be mixed in a quadratic
action that leads to a diagonalization of the operator in the BV action (24).

In the above parametrization the path integral for the BV action implies an
integration over the ISes (26). We define the measure on the space of such
superfields by the relation
\begin{eqnarray}
\int d\zeta \, e^{\zeta^2}& =& 1				    %(34)
\end{eqnarray}
Note that the expansions (30--33) have the general structure
\begin{eqnarray}
\Phi(k,l)& =& \zeta_0(k,l) + \sum_I \zeta_I(k_I,l_I)		  %(35)
\end{eqnarray}
with $ k_I+l_I < k+l$. One can show that for an arbitrary operator $A$
of the form
\begin{eqnarray}
& \prod\limits_i (Q-q_i\mu\bar{\mu})			   &	 %(36)
\end{eqnarray}
the relation holds (under the integration over the superspace):
\begin{eqnarray}
\Phi A \Phi& =& \zeta_0 A \zeta_0 + \sum_I \zeta_I B_I A \zeta_I  %(37)
\end{eqnarray}
where $ B_I $ have the similar form (36). Eq.(37) enables us to express
Gaussian integrals over the ISes (26)
\begin{eqnarray}
\int d\zeta \, e^{\zeta A \zeta}&{}&			 \nonumber
\end{eqnarray}
in terms of the integrals over the unconstrained and chiral superfields
which are usually defined [27,31]. This can be done by induction, the first
step being the integration over chiral superfields. Then the integral over
$\zeta_0$ from (35) can be expressed via the integrals over the unconstrained
superfield $\Phi$ and the ISes $\zeta_I$ with lower total number of indices:
\begin{eqnarray}
\int d\zeta_0(k,l) \, e^{\zeta_0 A \zeta_0} =
\int d\Phi \, e^{\Phi A \Phi} \,
\left[\prod\limits_I\int d\zeta_I\,
e^{\zeta_I A \zeta_I} \right]^{-1}.		   & &%(38)
\end{eqnarray}
The Jacobian of the change (35) takes the following form:
\begin{eqnarray}
J = \left[ \prod\limits_I\int d\zeta_I\,
e^{\zeta_I B_I \zeta_I} \right]^{-1}&& %(39)
\end{eqnarray}
Note that owing to Eq.(39) we can consider the Jacobian for each IS
separately.

\vspace{0.5cm}
\noindent
{\bf 5.} Let us continue with a description of the contributions to the
effective action. Consider, for example, the sector of the ISes $ \rho(k,k),
\rho'(k,k)$, and $\zeta(k,k) $ from (30,31) being transformed with the
parameter $ z^0(k,k) $ from (40) for $ k=0,\ldots,s-2$. To satisfy the
conditions (23) one should pass to the gauge invariant combination
$\zeta'(k,k)$,
\begin{eqnarray}
\zeta(k,k)& \rightarrow& \zeta'(k,k) = \zeta(k,k)          \nonumber\\
&&\mbox{}+\frac{s+k+1}s [Q-(s(s+1)+k(k+2))\mu\bar{\mu}]
\left( \frac i2 \rho(k,k) + \rho'(k,k) \right)                   %(40)
\end{eqnarray}
the Jacobian of the change (40) being unit. One can derive from
Eqs.(37,39) at $A=1$, that the relevant contributions to the Jacobian
of the change (21,22), $k=0$ and to the action $\phi^a \tilde{s}_{ab}
\phi^b$ in (24) are determined by the relations
\begin{eqnarray}
&H(s,s)^2 \;\sim\; \sum\limits_k \rho(k,k)\, \Omega(s-k,k,k)\, \rho(k,k) +
\sum\limits_k \rho'(k,k)\, \Omega(s-k,k,k)\, \rho'(k,k)           &  \\ %(41)
&\Psi(s-1,s-2)^2\; \sim \;\sum\limits_k \zeta(k,k)\, \Omega(s-k-1,k,k)\,
\zeta(k,k)							&\\%(42)
&(L^0(s,s-1))^2 \;\sim\; \sum\limits_k z^0(k,k)\, \Omega(s-k,k,k)\,
z^0(k,k)						   & \\ %(43)
&S^{\|}\; \sim\; \sum\limits_k
\zeta'(k,k)\,\Omega(s-k-1,k,k)\,\zeta'(k,k)&\\%(44)
&\Omega(m,k,l)\; =\; \prod\limits^m_{j=1}
\left[ Q - \frac{\mu\bar{\mu}}2 ((j+1)(j+k+l)-(k+l)(k+l+1)) \right] &%(45)
\end{eqnarray}
where only the sector at issue is extracted and numerical factors are
omitted. As the contributions from (43,44) enter with the sign
opposite to that for (41,42), the overall cancelation occurs. Then the
transformation law for the ISes $\rho$, $\rho'$ with the parameter $z^0$
does not contain derivatives, so the correspondent block in
$\tilde{Z}_{\nu_0}^{\mu_0}$ from (24) has unit determinant. The
analogous cancelation takes place in the sector of the ISes
$\zeta'(k,k-1)$, $\zeta(k,k-1)$, $k=1,\ldots,s-1$, (38,39).

In fact, after all cancellations the remained contributions are given in our
scheme by the following sectors of Eq.(30) only. The gauge invariant IS
$\rho(s,s)$ enters the action (24) and the purely gauge IS $\zeta'(s,s-1)$
contributes to the Jacobian of the change (30).  The contributions prove to
be proportional, giving the summary result
\begin{eqnarray}
-\frac 12 {\rm Tr}_{\rho(s,s)} \ln
[Q-s(2s+3)\mu\bar{\mu}]& +& \frac 14 {\rm Tr}_{\zeta(s,s-1)} \ln
[Q-s(2s+3)\mu\bar{\mu}] 				%(46)
\end{eqnarray}
where ${\rm Tr}_{\rho(s,s)}$ and ${\rm Tr}_{\zeta(s,s-1)}$ denote the trace
of the operator in the space of IS $\rho(s,s)$ and $\zeta(s,s-1)$
respectively. The result is natural firstly because $Q=s(2s+3) \mu\bar{\mu}$
is exactly the value of the Casimir in the massless representation of the
super AdS algebra with the superspin $s+1/2$.  Secondly Eq.(46) gives
effectively the trace of the operator in the space of an irreducible
representation. This can be established by counting the dimensions of the
ISes $\rho(s,s)$ and $\zeta(s,s-1)$.  In terms of the dimension of a complex
scalar field, $\dim \, \zeta(k,l) = 4(k+l+2),\; \dim \, \rho(s,s) = 2(2s+1)$.
So Eq.(46) means the trace over the space with the dimension
\begin{eqnarray}
\dim \rho(s,s) - \frac 12 \dim \zeta(s,s-1)& =& 2		    %(47)
\end{eqnarray}
that is exactly the dimension of the massless representation of the
AdS superalgebra [30].

All the contributions from the rest of ghost sector mutually cancel.  In some
more detail the ISes $a$, $\A$ (33) and $m$, $\mu$ (32,33) are purely gauge
superfields being transformed under (8,9,13--15) through the parameters $z$,
$\phi$ (32) and certain linear combinations $l$, $\lambda$ of the parameters
$m$, $\mu$ (32) and $n$, $\nu$ (33) without derivatives.  The Jacobian of the
change $\{n,\nu\} \to \{l,\lambda\}$ is unit. The Jacobian of the change
(33,34) corresponding to the ISes $a$, $\A$, $m$, $\mu$ is cancelled by that
corresponding to the ISes $z$, $\phi$, $n$, $\nu$. This ensures that every
ghost stage in (22,24) gives zero contribution to the effective action. Thus
even if the theory had the infinite stage of redusibility the ghost
contribution in the effective action would be equal to zero.

The latter phenomenon works for the transversal formulation (5) with
the gauge structure (7--11) of the infinite stage of reducibility and
enables us to quantize this formulation. It turns out that after a
series of cancellations the remained contributions are the same (46).
This proves the quantum equivalence of the formulations (5,6) which
are classically dual to each other [1--3]. Since we have used the
expression (3) for the variable $\Gamma(s-1,s-1)$, the considered
quantization admits the flat limit.

Thus we have accomplished the lagrangian quantization of the known
superfield formulations of massless theories of an arbitrary
half-integer superspin. The result (46) has the sense of the trace in
the physical subspace of the logarithm of the massless Casimir
$Q-s(2s+3)\mu\bar{\mu}$. The reducibility iterations
in the ghost sector gives zero contributions no matter is the stage
finite or not. Hence the theories of integer superspins can be
quantized in similar fashion.

\vspace{0.5cm}
\noindent
{\bf Acknowledgements:} I.L.B. and A.G.S. are grateful to S.J. Gates for
interesting discussions. The work of I.L.B. was supported in part by Grant
RI1000 from the International Science Foundation and the Russian Foundation
for Fundamental Research (project No 94--02--03234). The work of S.M.K. and
A.G.S. was supported in part by Grant No N2I000 from the International
Science Foundation and Grant No INTAS-93-2058 from the European Community.

\end{document}